\documentstyle[12pt,blackdvi,epsfig]{article}

\def \to {\rightarrow}

\def \beq {\begin{equation}}
\def \eeq {\end{equation}}

\def \ba {\begin{eqnarray}}
\def \ea {\end{eqnarray}}

\def \jpsi {J/\psi}
\def \twopi {\pi^+ \pi^-}

\def \smatr {\langle f | S | i \rangle}

\begin{document}
\begin{flushright}
AS-ITP-2001-005
\end{flushright}
\vskip 10mm
\begin{center}
{\LARGE Exclusive decay of $1^{- -}$ heavy quarkonium into photon
and two pions} \\
\vskip 10mm
J. P. Ma   \\
{\small {\it Institute of Theoretical Physics , Academia
Sinica, Beijing 100080, China }} \\
~~~ \\
Jia-Sheng Xu \\
{\small {\it China Center of Advance Science and Technology
(World Laboratory), Beijing 100080, China }} \\
{\small {\it and Institute of Theoretical Physics , Academia
Sinica, Beijing 100080, China }}
\end{center}

\begin{abstract}
We study the exclusive decay of $1^{--}$ heavy quarkonium into one
photon and two pions in the kinematic region, where
the two-pion system has a invariant mass which is much smaller
than the mass of heavy quarkonium.
Neglecting effects suppressed by the inverse
of the heavy quark mass, the decay amplitude can be factorized, in which
the nonperturbative effect related to heavy quarkonium is represented by a
non-relativistic QCD matrix element, and that related to the two
pions is
represented by a distribution
amplitude of two gluons in the isoscalar pion pair.
By taking the asymptotic form for
the distribution amplitude and by using chiral perturbative theory
we are able to obtain
numerical predictions for the decay.
Numerical results show that the decay of $\jpsi$ can be
observed at BEPC and at CESR.
Experiment observation of this process in this kinematic region at
BEPC and CESR  can provide information about
how gluons are converted into the two pions and may
supply  a unique approach to  study
$ I = 0$ s-wave $\pi \pi$ scattering.

\end{abstract}

\vfill\eject\pagestyle{plain}\setcounter{page}{1}


Recently it has been proposed to study production of two pions in
exclusive processes like $\gamma+\gamma^*\to \pi+\pi$\cite{diehl,D2,KMP}
and like $\gamma^* +h \to h +\pi+\pi$\cite{PL1,L1}.
These processes will enable us to
study how quarks and gluons, which are fundamental dynamical
freedoms of QCD, are transmitted into the two pions. In the
kinematic region, where the two pions have a much smaller
invariant mass than the virtuality of the virtual photon, the
scattering amplitudes of the processes take a factorized form. In
this factorized form the transition of partons into the two pions
are described by distribution amplitudes of partons in the two
pions, which are defined with twist-2 operators.
For $\gamma+\gamma^*\to \pi+\pi$,
at the tree-level,  only the distribution amplitude of quark appears
in the scattering amplitude,  the distribution amplitude of
gluon appears at loop levels or through evolution of distribution
amplitudes\cite{diehl,D2,KMP}, while for
$\gamma^* +h \to h +\pi+\pi$, at the tree-level, the distribution
amplitude of quark as well as that of gluon contribute to
the scattering amplitude, and the produced charged pion pair is
in isospin $I=0$ or $I=1$ states\cite{PL1,L1}.
In all of the above processes,
the produced system of two charged pions  will be dominantly
in an isospin $I\ne 0$ state.  This may make it difficult to
extract the distribution
amplitude of gluon from experimental data, because the two pions
produced through gluon conversion are in a $I=0$ state. In
this letter we propose to study the radiative decay of $\jpsi$
into two pions, where the distribution amplitude of gluon appears
at the tree-level and the produced two pions are dominantly in a
$I=0$ state. This makes the extraction of gluon content of a
two-pion system relatively easier in experiment.  Beside this
the decay also provides a possibility to study
$ I = 0$ s-wave $\pi \pi$ scattering.
The decay can be studied with the data sample of
$5\times 10^7$ $\jpsi$'s whose
collection will be completed with BES at the end of this year.

\par
We consider the decay in the kinematic region where the two pions
have a invariant mass which is much smaller than the mass of
$\jpsi$. In the limit of $m_c \to\infty$ the decay amplitude can
be factorized, where the nonperturbative effect of $\jpsi$ can be
presented by a NRQCD matrix element, while the nonperturbative
effect of the two pions is represented by the same distribution
amplitude of gluon appearing in the exclusive processes mentioned
above. Corrections to this limit may systematically be added. With
the model for the distribution amplitude of gluon, developed in
\cite{KMP,L1},
we obtain numerical predictions for the decay in the heavy mass
limit, and they indicate that the decay is observable at BES.

\par
We study the exclusive decay in the rest frame of $\jpsi$:
\begin{equation}
 \jpsi (P) \to \gamma(k_\gamma) +\pi^+(k_+) +\pi^-(k_-).
\end{equation}
The momenta are indicated in the brackets. We denote $k=k_++k_-$ and
$m^2_{\pi\pi}=k^2$.
At leading order of QED, the S-matrix element for the decay is
\begin{equation}
\smatr =  - i Q_c ~e \int d^4 z\langle
\gamma \twopi | A_{\mu} (z) \bar{c} (z) \gamma^{\mu} c(z) | \jpsi
\rangle ,
\end{equation}
where $Q_c$ is the electric charge of c-quark in unit of $e$,
$c(z)$ is the Dirac field for c-quark. At leading order of QCD,
two gluons are emitted by the c- or $\bar{c}$-quark, and these two
gluons will be transmitted into the two pions. Using Wick theorem
we obtain:
\ba
\smatr &=& \frac{i}{2} ~\frac{1}{2} \delta^{a b}
Q_c ~e ~g_s^2 ~\varepsilon_{\mu}^{*} (\gamma) \int d^4 z ~d^4 y
~d^4 x ~d^4 x_1 ~d^4 y_1
~e^{i (k_{\gamma}\cdot z + k_2\cdot y) }  \nonumber \\
&& \times ~\langle 0 | \bar{c}_j (x_1) c_i (y_1) | \jpsi \rangle \
 ~\langle \twopi |  G_{\mu_1}^{a} (x) G_{\nu_1}^{b} (0) |0
\rangle
\nonumber \\
&& \times ~[ \delta^{4} (x - x_1) \delta^{4} (z - y_1)
\gamma^{\mu_1}  S_F (x-y) \gamma^{\nu_1} S_F (y-z) \gamma^{\mu} +
\cdots ],
\ea
where $k_2$ is the momentum of one of emitted
gluons, $S_F (x - y)$ is the Feynman propagator of c-quark, the
dots in the square bracket denotes other five terms. In the limit
of $m_c \to\infty$, a $c$- or $\bar c$-quark moves with a small
velocity $v$, this fact enables us to describe nonperturbative
effect related to $\jpsi$ by NRQCD, in which a systematic
expansion in $v$ is employed. This expansion is extensively used
for inclusive decays\cite{nrqcd}. This expansion can also be used
here for the matrix $\langle 0 | \bar{c}_j (x_1) c_i (y_1) | \jpsi
\rangle$ , in which we expand the Dirac field $c(x)$ with
corresponding NRQCD fields. With the expansion we obtain:
\begin{equation}
\label{nrqcdmatr} \langle 0 |\bar{c}_j (x_1) c_i (y_1)| \jpsi
\rangle ~= - \frac{1}{6} ~(P_+ ~\gamma^l ~P_-)_{ij} ~\langle 0 |
\chi^{\dagger} \sigma^l \psi | \jpsi \rangle ~ e^{- i p\cdot (x_1
+ y_1)} + {\rm O} (v^2),
\end{equation}
where $\chi^{\dagger} (\psi)$ is the NRQCD field for $\bar{c} (c)$
quark, $\sigma^l  (l = 1, 2, 3)$ is the Pauli matrix, and \ba
P_{\pm} &=& \frac{1}{2} ~( 1 \pm \gamma^0 ) \nonumber \\
p &=& ( m_c, 0, 0, 0 ). \ea The matrix $\langle 0 | \chi^{\dagger}
\sigma^l \psi | \jpsi \rangle $ is proportional to the
polarization vector $\varepsilon ^l  (\jpsi)$  at the considered
order. In this paper, we neglect the contribution from higher
orders in $v$, the momentum of $\jpsi$ is then approximated by $2
p$. It should be noted that the effect of higher order in $v$ may
be added.

\par
\begin{figure}[hbt]
\vspace*{1mm}
\centerline{\epsfxsize=7cm \epsffile{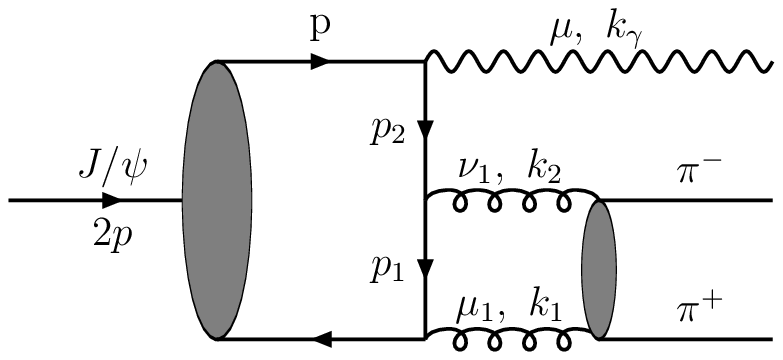}}
\vspace*{5mm}
\caption{ One of the Feynman diagrams for the exclusive decay of
$\jpsi$ into one photon and two pions.
\label{Feynman-dg} }
\end{figure}

Using Eq. (\ref{nrqcdmatr}) we can write the S-matrix element as
\begin{eqnarray}
\label{smatrix} \smatr &=& \frac{- i}{24} Q_c ~e ~g_s^2 ~(2
\pi)^4 ~\delta^{4} (2 p - k -k_{\gamma}) ~\varepsilon_{\mu}^{*}
(\gamma) ~\langle 0 | \chi^{\dagger} \sigma^l \psi | \jpsi \rangle
\nonumber\\
&& \int \frac{d^4 k_1}{(2 \pi)^4}  H^{l \mu \mu_1 \nu_1} (p, k, k_1)
~\Gamma_{\mu_1 \nu_1} (k, k_1) ,
\end{eqnarray}
and
\beq
\Gamma^{\mu \nu}
(k, k_1) = \int d^4 x ~e^{- i k_1 x} ~\langle \twopi |
G^{a,\mu} (x) G^{a,\nu} (0)| 0 \rangle,
\eeq
in which
$H^{l \mu \mu_1\nu_1} (p, k, k_1)$ is a perturbative function,
which is the amplitude for a $c\bar c$ pair emitting a photon and
two gluons.
The contributions in Eq.(\ref{smatrix}) can be
represented by Feyman diagrams. One of them is given in
Fig.\ref{Feynman-dg}, where the kinematic variables are also
indicated. The nonperturbative object $\Gamma^{\mu \nu}(k, k_1)$
describes how two gluons are converted into $2\pi$.

\par
In the heavy quark limit the two-pion system will have
a large momentum $k$, we
take a coordinate system in which the direction of $\bf k$ is
the $z$-direction. For convenience we will work
in the light-cone coordinate system, in which the components of $k$ are
given by

\begin{equation}
 k^\mu =(k^+, k^-,{\bf 0}), \ \ k^+ =(k^0+k^3)/\sqrt{2},\ \
 k^- =(k^0-k^3)/\sqrt{2}.
\end{equation}
In the light-cone coordinate system we introduce two light cone vectors
and a tensor:
\begin{eqnarray}
n^{\mu} &=& (0, 1, 0, 0),  \ \
l^{\mu} = (1, 0, 0, 0),\nonumber \\
d_T^{\mu\nu} &=& g^{\mu\nu} -n^\mu l^\nu -n^\nu l^\mu,
\end{eqnarray}
and we take the gauge
\begin{equation}
 n\cdot G(x) =0 .
\end{equation}

\par
The $x$-dependence of the matrix element in Eq.(7) is controlled by different
scales. The $x^-$-dependence is controlled by $k^+$, which is large, while
the $x^+-$ and ${\bf x_T}$-dependence are controlled by the scale $k^-$ and
$\Lambda_{QCD}$, which are small in comparison with $k^+$. With this observation
we can expand the matrix element in $x^-$ and in ${\bf x_T}$. At the leading order
only twist-2 operators contributes to the matrix element. We will neglect higher
orders in the expansion, i.e., we only keep contributions of twist-2 operators.
Then we obtain:

\begin{equation}
\label{gammunu}
\Gamma^{\mu \nu} (k, k_1) = (2 \pi)^4 ~\delta ( k_1^- )
~\delta^2 ( k_{1T} )~\frac{1}{k^+} ~\frac{1}{x_1 (1 - x_1)} \nonumber \\
\left[ \frac{1}{2} ~d_T^{\mu \nu}
~\Phi^{G} ( x_1, \zeta, m_{\pi \pi} )  \right] ,
\end{equation}
with
\begin{eqnarray}
\Phi^{G} ( x_1, \zeta, m_{\pi \pi} )&=& \frac{1}{k^+}
~\int \frac{d x^-}{2 \pi} ~e^{ - i k_1^+ x^-}
\times \langle \twopi |G^{a,+\mu} (x^-n) G^{a,+}_{\ \ \ \mu} (0)| 0 \rangle,
\nonumber \\
x_1 &=& \frac{k_1\cdot n}{k\cdot n}, \ \ \
\zeta = \frac{k_+\cdot n}{k\cdot n}.
\end{eqnarray}
$\Phi^{G} ( x_1, \zeta, m_{\pi \pi} )$ is the gluonic distribution
amplitude which describe how a pion pair is produced by two
collinear gluons, it represents a nonperturbative effect and can
only be calculated with nonperturbative methods or extracted from
experiment. As it stands, it is gauge invariant in the gauge
$n\cdot G(x) =0$. In other gauges we need to supply a Wilson line
operator to make it gauge invariance. With Eq.(11) it is
straightforward to obtain the $S$-matrix element at the tree-level
in our approximation:
\begin{eqnarray}
\label{smatrix-f}
\smatr &=& \frac{- i}{12m_c^2} Q_c ~e ~g_s^2 ~(2 \pi)^4 ~\delta^{4}
(2 p - k -k_{\gamma})
\varepsilon^{*\ell} (\gamma)  \langle 0 | \chi^{\dagger}
   \sigma^\ell \psi | \jpsi \rangle  \nonumber \\
   && \int_0^1  d x_1 \frac{1}{x_1 (1 - x_1) }\cdot
 \frac{ 8 m_c^2 (1 - x_1) - m_{\pi\pi}^2 (1 - 2 x_1) }
          { 4 m_c^2 (1 - x_1) - m_{\pi\pi}^2 (1 - 2 x_1) }
 \cdot \Phi^{G} ( x_1, \zeta, m_{\pi \pi} ).
\end{eqnarray}
In the calculation we have kept the mass $m_{\pi\pi}$.
Usually the effect of $m_{\pi\pi}$ should be combined with
effects of twist-4 operators as a correction to the above
result. We have checked that striping the $m_{\pi\pi}$
terms changes the results by less than $5\%$.
In Eq.(13) the nonperturbative effect related to
$\jpsi$ and that to the two-pion system are separated, the former
is represented by a NRQCD matrix element, while the later is
represented by the matrix element of gluonic operators, which is
defined in Eq.(12). The NRQCD matrix element is related to the
wave-function of $\jpsi$ in potential models and can be estimated
with these models. It can also be calculated with lattice QCD or
extracted from experiment. Little is known numerically about the
matrix element of gluonic operators, i.e., the function $\Phi^{G}
( x_1, \zeta, m_{\pi \pi} )$. A detailed study of properties
of $\Phi^{G}(x_1,\zeta,m_{\pi \pi})$ can be found in \cite{D2,KMP,L1}.
For our numerical prediction we will use their results for $\Phi^{G}
( x_1, \zeta, m_{\pi \pi} )$, in which asymptotic form of
$\Phi^{G} ( x_1, \zeta, m_{\pi \pi} )$ is taken as an Ansatz for
$\Phi^{G} ( x_1, \zeta, m_{\pi \pi} )$. It should be noted that
the renormalization scale $\mu$ should be taken as $2 m_c$ in our case.
Because it is
not very large, the actual shape of $\Phi^{G} ( x_1, \zeta, m_{\pi
\pi} )$ may look dramatically different than that of the
asymptotic form. Keeping this in mind we take the form $\Phi^{G} (
x_1, \zeta, m_{\pi \pi} )$ as that given in \cite{L1}:
\begin{equation}
\Phi^{G} ( x_1, \zeta, m_{\pi \pi} )= - 60 ~M_2^G ~x_1^2 ~(1 - x_1)^2
\left[ \frac{3 C - \beta^2}{12} ~
f_0 ( m_{\pi\pi}) ~ P_0 (\cos \theta_c) \right. \nonumber \\
\left. - \frac{\beta^2}{6} ~f_2 (m_{\pi\pi}) ~P_2 (\cos \theta_c)
\right] ~,
\end{equation}

\begin{figure}[htb]
\begin{center}
\vskip -5mm
\epsfig{file=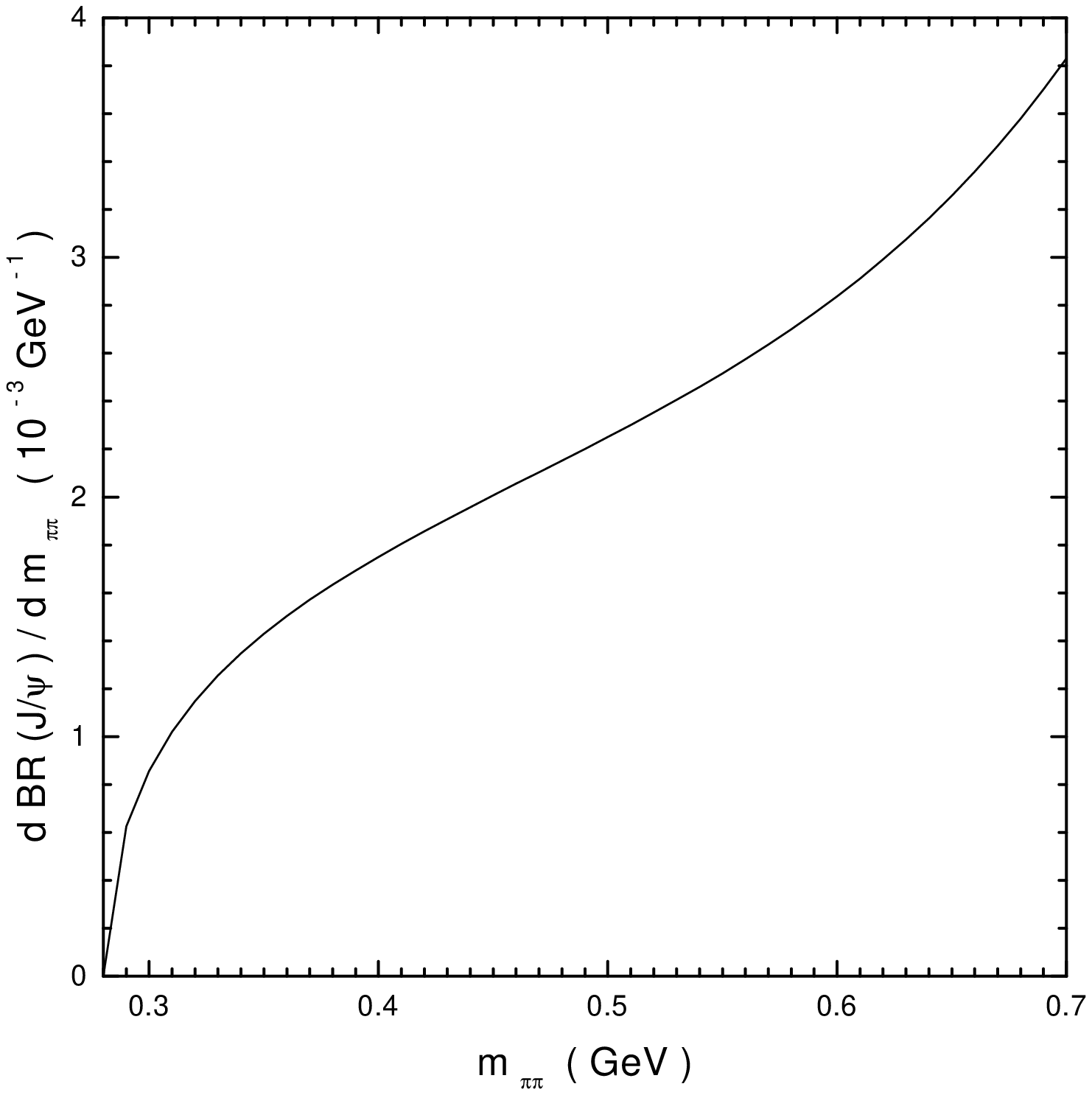,width=8cm}
\end{center}
\vskip -55mm
\caption{The differential decay branching ratio
$d ~{\rm BR} (\jpsi)/dm_{\pi\pi}$ as a function of $m_{\pi\pi}$
in unit of $ 10^{-3} {\rm GeV}^{-1}$.
\label{mpipi-ds}}
\end{figure}

\noindent
where $\theta_c$ is the scattering angle of $\pi^+$ and
$\beta$ is the velocity of $\pi^+$ in the center of mass system of
the two pions, they are related to $\zeta$ and $m_{\pi\pi}$ by
\beq \beta ~\cos \theta_c  = 2 ~\zeta - 1, ~~\beta = \sqrt{1 -
\frac{4 ~m_{\pi}^{2} }{m_{\pi \pi}^{2}}} ~.
\eeq
$C$ is a constant
and takes $C = 1 + b ~m_{\pi}^{2} + O (m_{\pi}^{4}) $ with $b
\simeq - 1.7 {\rm GeV}^{- 2}$ \cite{KMP,L1}, $M_2^G$ is the
momentum fraction carried by gluons in the pion, its asymptotic
value  is
\beq M_2^G = \frac{ 4 ~C_F}{ N_f + 4 ~C_F}.
\eeq

\begin{figure}[htb]
\begin{center}
\vspace*{-5mm}
\epsfig{file=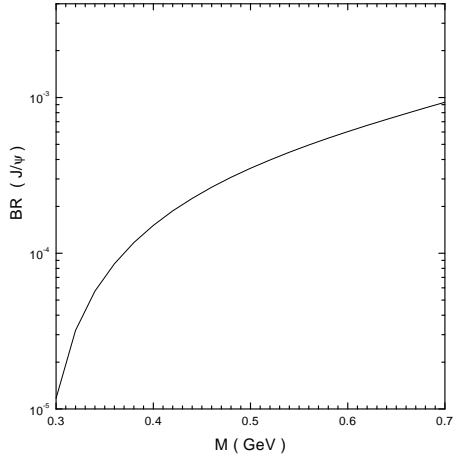,width=8cm}
\end{center}
\vskip -55mm
\caption{The decay branching ratio of
$\jpsi \to \gamma + \pi^+ \pi^-$ as a function of $M$, the upper cut of $m_{\pi\pi}$.
\label{maxmpipi-ds}}
\end{figure}

\noindent
$f_0 (m_{\pi\pi})$ and $f_2 (m_{\pi\pi})$  are the Omn\`{e}s
functions for ${\rm I} = 0$ S- and D-wave $ \pi \pi$ scattering,
respectively. The Omn\`{e}s function  $f_2 (m_{\pi\pi})$ is
dominated by the $f_2 (1270)$ resonance resulting a peak at
$m_{\pi\pi} = 1.275 {\rm GeV}$, while the Omn\`{e}s function  $f_0
(m_{\pi\pi})$ in the relevant $m_{\pi\pi}$ region we studied
($m_{\pi\pi} \leq 0.70 {\rm GeV}$) can be calculated by the chiral
perturbative theory, the result is \cite{donoghue}
\begin{equation}
f_0(m_{\pi\pi})= 1 + \frac{m_{\pi\pi}^{2}}{192 ~\pi^2 ~f_{\pi}^{2}}
 + \frac{2 m_{\pi\pi}^{2} - m_{\pi}^{2}}{32 ~\pi^2 ~f_{\pi}^{2}}
\left[ \beta ~\ln~\left(\frac{1 - \beta}{1 + \beta}
\right) + 2 + i \pi \beta \right] ,
\end{equation}
where $f_{\pi}=93{\rm MeV}$is the pion decay constant.

\begin{figure}[htb]
\begin{center}
\vskip -8mm
\epsfig{file=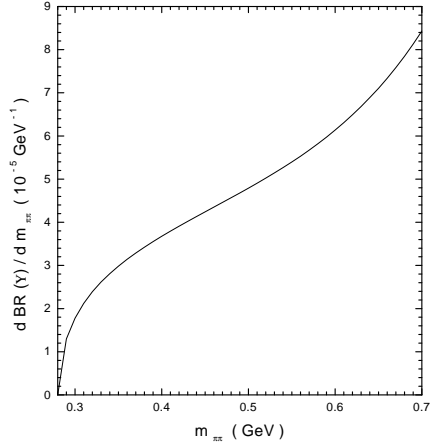,width=8cm}
\end{center}
\vskip -55mm
\caption{The differential decay branching ratio
$d ~{\rm BR} (\Upsilon) /dm_{\pi\pi}$ as a function of $m_{\pi\pi}$
in unit of $ 10^{-5} {\rm GeV}^{-1}$.
\label{bmpipi-ds}}
\end{figure}

\par
With these results we are able to predict the shape
of the differential decay branching ratio
$d ~{\rm BR} (\jpsi)/dm_{\pi\pi}$
numerically where we use
the leptonic decay of $\jpsi$ to determine the NRQCD
matrix element. The result is presented in Fig.2, where
we take the parameters: $m_c=1.5$GeV and $\alpha_s(2 m_c)=0.31$.
We also calculate the decay branching ratio, where we use a upper cut
$M$ for $m_{\pi\pi}$, hence the decay branching width is a function
of $M$:
\begin{equation}
 {\rm BR} (\jpsi)  = \int_{2m_\pi}^M d m_{\pi\pi}
   \frac{d ~{\rm BR} (\jpsi)}{dm_{\pi\pi}}.
\end{equation}
The numerical result for this  decay ratio is given in Fig.3.
For $M > 0.38 {\rm GeV}$, we can read from Fig.3 that
the branching ratio is larger than $10^{-4}$,
for $M=0.70 {\rm GeV}$
the branching  ratio is $ 9.3 \times 10^{ - 4}$.
Therefore it is likely that the decay mode
can be measured using the data collected at BES.
In the kinematic region we consider the dominant contribution to
the differential decay width is from $f_0(m_{\pi\pi})$. The contribution
from $f_2(m_{\pi\pi})$ is less than $10\%$. This gives
a possibility to  study
$ I = 0$ s-wave $\pi \pi$ scattering in the decay.
Similar results for $\Upsilon$-decay are also obtained, which are
presented in Fig.4 and in Fig. 5.
For  $m_{\pi\pi} \leq 0.70 {\rm GeV}$, the branching ratio is
$2.0 \times 10^{-5}$.

\begin{figure}[htb]
\begin{center}
\vskip -5mm
\epsfig{file=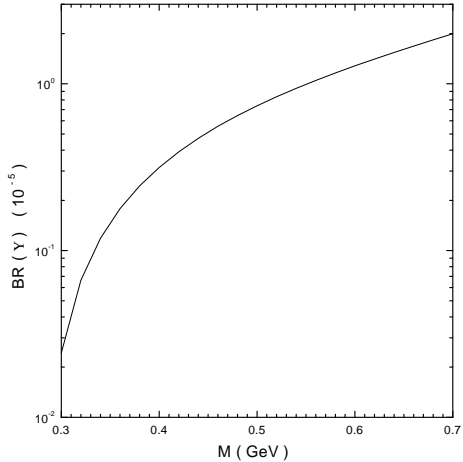,width=8cm}
\end{center}
\vskip -55mm
\caption{The decay branching ratio of $\Upsilon \to \gamma + \pi^+ \pi^-$
as a function of $M$, the upper cut of $m_{\pi\pi}$ in unit of $10^{-5}$.
\label{bmaxmpipi-ds}}
\end{figure}

\par
It should be noted that the two-pion system produced in the decay is
in a S-wave state in our approximation. At twist-2 level, the produced system
can also be in a D-wave state, the corresponding tensor distribution
amplitude can be found in \cite{KMP}. However the decay amplitude
for a D-wave state is kinematically suppressed, because it is
proportional to $(k_+-k_-)^\mu (k_+-k_-)^\nu$ and the relative momentum
$(k_+-k_-)$ of the two pions is a small quantity in the kinematic region
considered here.
If one replaces the two-pions system by the
tensor meson $f_2(1270)$, one also finds that the decay amplitude
for $f_2$ with the helicity $\pm 2$ is kinematically suppressed\cite{Ma}.
Although the effect of the D-wave state
is suppressed, it can lead to some nonzero azimuthal asymmetries which
may be observable in experiment.
\par
To summarize: In this letter we have studied the exclusive decay of $1^{--}$
heavy quarkonium into one
photon and two pions.  At leading twist, the S-matrix of the decay takes
a factorized form, the nonperturbative
effect related to heavy quarkonium and that related to the light
hadrons can be separated, the former is represented by a
non-relativistic QCD matrix element, while the later is
represented by the two-gluon to isoscalar two-pion distribution
amplitude. By taking the asymptotic form for
the distribution amplitude and by using chiral perturbative theory
we are able to obtain numerical predictions for the decay.
For $\jpsi$ , the branch ratio for $m_{\pi\pi} \leq 0.70 {\rm GeV}$
is $ 9.3 \times 10^{ - 4}$, while for $\Upsilon$,
the branch ratio for $m_{\pi\pi} \leq 0.70 {\rm GeV}$ is
$2.0 \times 10^{-5}$. The decay of $\jpsi$ can be studied
at BES and the study  can provide a possibility to  study
$ I = 0$ s-wave $\pi \pi$ scattering.

\newpage
\begin{center}
{\bf\large Acknowledgments}
\end{center}

The work of J. P. Ma is supported  by National Science Foundation of P. R. China and by the
Hundred Young Scientist Program of Academia Sinica of P. R. China,
the work of J. S. Xu is supported by the Postdoctoral Foundation of P. R. China and  by
The K. C. Wong Education Foundation, Hong Kong.

\end{document}